\documentclass[reqno,10pt]{article}
\usepackage{graphicx}
\usepackage{amsmath}
\usepackage{amssymb}
\usepackage{amsthm}
\usepackage{enumitem}
\usepackage{bbm}
\usepackage[l2tabu,orthodox]{nag}
\usepackage[all,error]{onlyamsmath}
\usepackage[strict]{csquotes}
\usepackage{url}
\usepackage{color}

\theoremstyle{definition}

\numberwithin{equation}{section}
\numberwithin{theorem}{section}

\newenvironment{OMabstract}{\noindent\textbf{Abstract.} }{\medskip}
\newenvironment{OMsubjclass}{\noindent\textbf{Mathematics Subject Classification (2020):} }{\medskip}
\newenvironment{OMkeywords}{\noindent\textbf{Keywords:}  }{\medskip}

\begin{document}

\author{Mats-Erik Pistol$^{1}$, Vyacheslav Pivovarchik$^{2}$}

\title{Cospectral trees indistinguishable by scattering }
\maketitle


\begin{OMabstract}
 Let $v_1$ and $v_2$ be two distinct vertices of a tree $T_0$. Let $\phi_N^{(i)}$ ($i=1,2$)  be the characteristic functions of the Sturm-Liouville problem on $T_0$ rooted at $v_i$ with Neumann conditions at the root  and let $\phi_D^{(i)}$ ($i=1,2$)  be the characteristic functions of the Sturm-Liouville problem on $T_0$  with Dirichlet  conditions at the root.   
We prove that if attaching any tree  to $T_0$ at the vertices $v_1$ and $v_2$ leads to cospectral trees and $d(v_1)=d(v_2)$ then $\phi_N(\lambda)^{(1)}\equiv \phi_N(\lambda)^{(2)}$ and $\phi_D(\lambda)^{(1)}\equiv \phi_D(\lambda)^{(1)}$  (which means that the scattering is the same at $v_1$ and $v_2$).    
\end{OMabstract}


\begin{OMkeywords}
     Sturm-Liouville equation, eigenvalue, spectrum, equilateral tree,  Dirichlet boundary condition, Neumann boundary condition, root.
\end{OMkeywords}

\begin{OMsubjclass}
    34B45, 34B240, 34L20
\end{OMsubjclass}


\section{Introduction}  

In quantum graph theory, i.e. in the theory of quantum mechanical equations considered on metric graph domains, the problem of recovering the shape of a graph was considered  in \cite{Be}, and \cite{GS}. 
It was shown in \cite{GS} that if the lengths of the edges are rationally independent then the spectrum of the spectral Sturm-Liouville problem on a graph with standard  boundary conditions (Neuman-Kirchhoff at interior vertices and Neumann at the pendant vertices) uniquely determines the shape of this graph. 

In \cite{Be}, it was shown that if one considers graphs having edges whose quotient is a rational number, then there exist co-spectral quantum graphs. 

In \cite{KN} it was shown that if we have the Neumann problem with zero potential  then the finite interval does not have any cospectral partners.
In \cite{CP} it was shown that if the graph is a simple connected equilateral graph where the number of vertices less or equal 5 and the potentials on the edges are real $L_2$ functions then the spectrum of the Sturm-Liouville problem with standard conditions at the vertices uniquely determines the shape of the graph. For trees the minimal number of vertices in a cospectral pair is 9 (see \cite{Pist} and \cite{CP1}). If the number of vertices doesn't exceed 8 then the characteristic polynomial determines the graph \cite{CP}. 

If one  spectrum does not determine uniquely the shape of a graph then we put the question whether two spectra do it, where the two spectra are obtained under different boundary conditions.

In \cite{P0} it was shown how to find the shape of a tree using  two spectra:  the spectrum of the  Neumann problem and the spectrum of the Dirichlet problem, i.e. the problem in which the Dirichlet condition is imposed at particular vertex called the root. This method works for arbitrarily large number of vertices. If the solution is not unique we can find all the solutions. 
 
In \cite{MuP} it was shown how to find the shape of a graph using the S-function and the eigenvalues of the scattering problem on a graph which consists of an equilateral compact subgraph with a lead attached to it at a vertex which we call the root. The potential on the lead was assumed to be zero identically and therefore the Jost-function can be expressed via the characteristic functions of the Dirichlet and Neumann problems. Thus this scattering inverse problem and the spectral inverse problem by two spectra have the same given data. 
However, it was found in \cite{Pist} that there are pairs of graphs that are cospectral also under very general boundary conditions.

Attaching a lead at different interior vertices may lead to different scattering data ($S$-functions and eigenvalues). However, an example of a tree with the same $M$-function for two different vertices was found in \cite{Pist} (which means that the $S$-function at these vertices and eigenvalues are the same if we attach any compact graph at these vertices).

In present paper we show that if we have a tree such that attaching an interval at two different vertices give cospectral graphs then the $M$-function at these vertices must be the same. It has previously been shown that attaching any compact graph to a graph having two vertices with the same $M$-function results in cospectral graphs.


\section{Statement of the problem and auxiliary results}

\vspace{1mm}

Let $T$ be a metric equilateral  tree with $p$ vertices and $g=p-1$ edges each of the length $l$. 

 We choose the vertex $v_0$ as the root and direct all  the edges  away from the root.  

Let us describe the {\it Neumann} spectral  problem on this tree. We consider  the Sturm-Liouville equations on the edges 
\begin{equation}
\label{2.1}
-y_j^{\prime\prime}+q_j(x)y_j=\lambda y_j, \ \  j=1,2,..., g 
\end{equation} 
where $q_j\in L_2(0,l)$ are real.

If an edge $e_j$ is incident with a pendant vertex which is not the root then
we impose the Dirichlet boundary conditions at the pendant vertex
\begin{equation}
\label{2.2}
 y_j(l)=0. 
\end{equation}

At each interior vertex, $v_i$, we impose the continuity conditions   
\begin{equation}
\label{2.3}
y_j(l)=y_k(0)
\end{equation}
for the incoming edge $e_j$ and for all $e_k$ outgoing from $v_i$ and the Kirchhoff's conditions
\begin{equation}
\label{2.4}
y'_j(l)=\mathop{\sum}\limits_k y_k'(0)
\end{equation}
where the sum is taken over all edges $e_k$ outgoing from $v_i$. 
At the root we have the continuity conditions

\begin{equation}
\label{2.5}
y_j(0)=y_k(0)
\end{equation}
for the outgoing from $v_i$ edges $e_j$ and  $e_k$ and the Kirchhoff's conditions
\begin{equation}
\label{2.6}
\mathop{\sum}\limits_k y_k'(0)=0
\end{equation}
where the sum is taken over all edges $e_k$ outgoing from $v_i$

Problem (\ref{2.1})--(\ref{2.6}) we will call the Neumann problem. Conditions (\ref{2.3}), (\ref{2.4}) and (\ref{2.4}), (\ref{2.5}) we call standard (or generalized Neumann). At pendant vertices standard conditions are Neumann.

Let us describe the corresponding Dirichlet problem. It consists of  equations (\ref{2.1})--(\ref{2.4}) and of equation
\begin{equation}
\label{2.7}
y_j(0)=0
\end{equation}
for all edges $e_j$ incident with the root. We illustrate the situations in Fig. 1.

 \begin{figure}[h]
 \begin{center}
   \includegraphics[scale= 0.6 ] {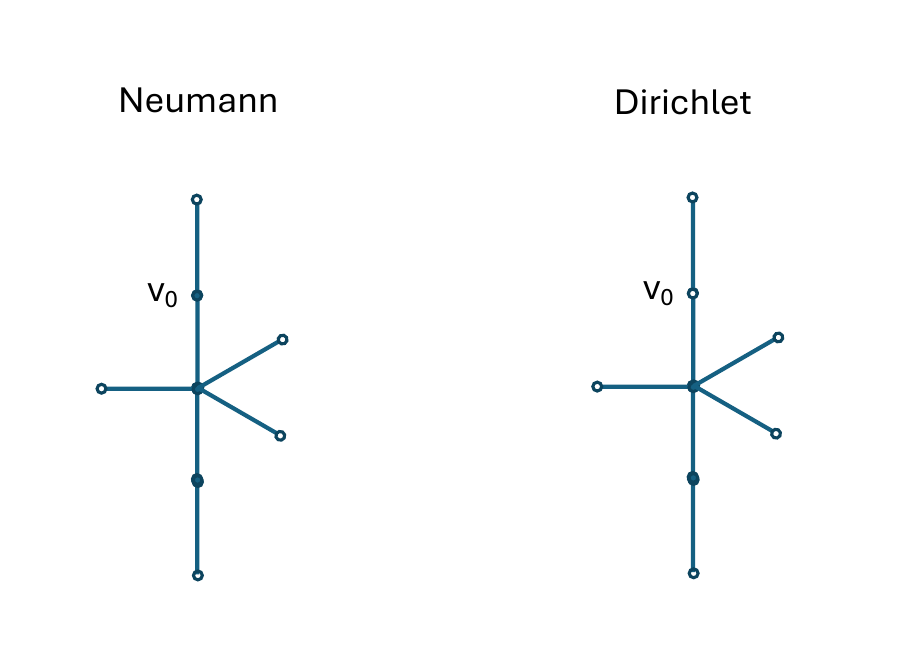}
  \end{center}
\caption{Open circles mean Dirichlet boundary conditions. Closed circles mean Neumann boundary conditions.}
\end{figure}

Denote by $s_j(\lambda,x)$ the solution of (\ref{2.1}) which satisfies the conditions $s_j(\lambda,0)=s'_j(\lambda,0)-1=0$ and by $c_j(\lambda,x)$ the solution which stisfies 
$c_j(\lambda,0)-1=c'_j(\lambda,0=0$.  Substituting 
\begin{equation}
\label{2.8}
y_j(\lambda,x)=A_js_j(\lambda,x)+B_jc(\lambda,x)
\end{equation} 
into equations (\ref{2.2})--(\ref{2.6}) we obtain a system of $2q$ homogeneous linear algebraic equations with unknowns $A_j$ and $B_j$. The determinant $\psi_N(\lambda)$ of the matrix of this system we call the characteristic function of the Neumann problem. The zeros of $\phi_N(\lambda)$ are the eigenvalues of problem (\ref{2.1})--(\ref{2.6}). Substituting (\ref{2.8}) into (\ref{2.2})--(\ref{2.4}), (\ref{2.7}) we again obtain a system of $2g$ homogeneous linear algebraic equations. The determinant $\psi_D(\lambda)$ of this system we call the characteristic function of the Dirichlet problem.

Let us assume that the root ${ v_0}$ is an interior vertex. We
divide our tree $T$ into two subtrees $T_1$ and $T_2$ having ${\bf
v}$ as the only common vertex. (We say that $T_1$ and $T_2$ are
{\it complementary subtrees} of $T$. Denote by $\psi^{(i)}_N(\lambda)$ the characteristic function of the Neumann problem on $T_i$ and by $\psi^{(i)}_D(\lambda)$ the characteristic function of the Dirichlet problem on $T_i$. 

The following theorem was proved in \cite{LP}.

{\bf Theorem 2.1}
Let the root ${\bf v}$ of a tree $T$ be an interior vertex. Let
  $T_1$ and $T_2$ be complementary subtrees of $T$.
 Then with the same orientation of the
 graph and the subgraphs edges described above,
\begin{equation}
\label{2.18}
\psi_N(z)=\psi_N^{(1)}(z)\psi_D^{(2)}(z)+\psi_D^{(1)}(z)\psi_N^{(2)}(z),
\end{equation}
\[
\psi_D(z)=\psi_D^{(1)}(z)\psi_D^{(2)}(z).
\]

\section{Main results}

Let $T_0$ and ${\tilde T}$ be equilateral trees with the lengths of the edges equal $l$, the same real potential on the edges $q(x)\in L_2(0,l)$ symmetric with respect to the midpoint of an edge ($q(l-x)\mathop{=}\limits^{a.e.}q(x)$) and is real. Let $v_1$ and $v_2$ be distinct interior vertices of $T_0$ of the degrees $d(v_1)$ and $d(v_2)$, respectively. Let $v_0$ be a  vertex of ${\tilde T}$ (the root of ${\tilde T}$). We attach ${\tilde T}$ to the  tree $T_0$ in such a way that $v_0$ coinciding with $v_i$ ($i=1,2$) is the only common vertex of ${\tilde T}$ and $T_0$. Thus we obtain two trees $T_1$ and $T_2$.

Then by Theorem 2.1
\begin{equation}
\label{3.1}
\phi_{i}(\lambda)=\tilde\phi_N(\lambda)\phi^{(i)}_{D}(\lambda)+\tilde{\phi_D}(\lambda)\phi^{(i)}_{N}(\lambda)
\end{equation}
where $\phi_i(\lambda)$  is the characteristic function of  problem (\ref{2.1})-(\ref{2.6}) on $T_i$ 
($i=1,2$), $\phi_N^{(1)}(\lambda)\equiv\phi_N^{(2)}(\lambda)$ is  the characteristic function   of (\ref{2.1})--(\ref{2.6})  on $T_0$ (the functions  $\phi_N^{(1)}(\lambda)$ and $\phi_N^{(2)}(\lambda)$ are the same because they correspond to the same (standard) conditions at $v_1$ and $v_2$), $\phi^{(i)}_{D}(\lambda)$ ($i=1,2$) is the Dirichlet characteristic function (characteristic function of problem (\ref{2.1})--(\ref{2.4}), (\ref{2.7}) on $T_0$ with the Dirichlet condition at the root  $v_i$, $\tilde{\phi}_{N}(\lambda)$ is  the Neumann characteristic functions on $\tilde{T}$ and $\tilde{\phi}_D(\lambda)$ is the Dirichlet characteristic function with the Dirichlet  condition at $v_0$.

The following result which follows from (\ref{3.1}) is known even for more general case (see \cite{KM}). 

{\bf Theorem 3.1} Suppose $\phi_{N}^{(1)}(\lambda)\equiv\phi_{N}^{(2)}(\lambda)$ and  $\phi_D^{(1)}(\lambda)\equiv\phi_D^{(2)}(\lambda)$. Then attaching a  tree ${\tilde T}$ by its root to the vertices $v_1$ and $v_2$ and imposing the same conditions at all the other vertices of ${\tilde T}$ we obtain co-spectral trees.


{\bf Lemma 3.2} Let the equilateral trees $T_1$ and $T_2$ (constructed by attaching a tree $\tilde{T}_0$ to interior vertices $v_1$ and $v_2$ of a tree $T_0$. Let the problems (\ref{2.1})--(\ref{2.4}), (\ref{2.7}) with the same real symmetric potential on the edges be co-spectral. Then  $\phi_{1}(\lambda)\equiv \phi_{2}(\lambda)$ if and only if $d(v_1)=d(v_2)$.

{\bf Proof} We use representations  \cite{MP}, Theorem 6.4.2:
\[
\phi_{i}=s^{p-g+r}(\lambda,l)P_i(c(\lambda,l)) \ \ (i=1,2)
\]
where  $p$ is the number of vertices, $g$ is the number of edges, $r$ is the number of pendant vertices with the Dirichlet conditions, $P_i(z)=det(-\lambda \hat{ D}_i+\hat{A}_i)$,
  $D_i =diag\{d(v_1), d(v_2),...,d(v_p)\}$ are the diagonal degree matrices, $A_i$ are the adjacency matrices, $-\lambda\hat{D}_i+\hat{A}_i$ are the prime submatrices of  $-\lambda D_i+A_i$, obtained by deleting the columns and rows corresponding to the pendant vertices.

 The trees $T_1$ and $T_2$ are co-spectral what means that the sets of zeros of $P_1(z)$ and $P_2(z)$ coincide.  The leading term of the polynomial $P_1$ is 
\[
(d(v_1)+d(v_0))d(v_2)\mathop{\prod}\limits_{i=3}^p d(v_i)z^p
\] 
while the leading term of the polynomial $P_2$ is 
\[
d(v_1)(d(v_2)+d(v_0))\mathop{\prod}\limits_{i=3}^p d(v_i)z^p.
\]
Thus,  $P_1(z)=CP_2(z)$
where $C=\frac{d(v_0)+d(v_1)}{d(v_0)+d(v_2)}$ and, therefore, $\phi_1(z)=C\phi_2(z)$. We conclude that $\phi_{1}(\lambda)\equiv \phi_{2}(\lambda)$ if and only if $d(v_1)=d(v_2)$.

{\bf Theorem 3.3} Let the trees $T_1$ and $T_2$ be co-spectral, i.e. the sets of zeros of $\phi_1(z)$ and $\phi_2(z)$ coincide and let $d(v_1)=d(v_2)$. Then $\phi_N^{(1)}(\lambda)\equiv\phi_N^{(2)}(\lambda)$ and $\phi_D^{(1)}(\lambda)\equiv\phi_D^{(2)}(\lambda)$. 

{\bf Proof} First, we conclude that $\phi_N^{(1)}(\lambda)\equiv\phi_N^{(2)}(\lambda)$ because $\phi_N^{(1)}(\lambda)$ and $\phi_N^{(2)}(\lambda)$ are the characteristic functions of the same problem, since conditions are standard at both vertices $v_1$ and $v_2$.

 By Lemma 3.2 we have $\phi_1(\lambda)\equiv\phi_2(\lambda)$. Due to (\ref{3.1}) we have 
\begin{equation}
\label{3.2}
\phi_{1}(\lambda)\equiv\phi_{2}(\lambda)=\tilde\phi_N(\lambda)\phi^{(1)}_{D}(\lambda)+\tilde{\phi_D}(\lambda)\phi^{(1)}_{N}(\lambda)=\tilde\phi_N(\lambda)\phi^{(2)}_{D}(\lambda)+\tilde{\phi_D}(\lambda)\phi_{N}^{(2)}(\lambda)
\end{equation}
and, consequently, 
\[
\phi_D^{(1)}(\lambda)=\frac{\phi_{1}(\lambda)-\tilde{\phi}_D(\lambda)\phi_N^{(1)}(\lambda)}{\tilde{\phi}_N(\lambda)}=\frac{\phi_{2}(\lambda)-\tilde{\phi}_D(\lambda)\phi_N^{(2)}(\lambda)}{\tilde{\phi}_N(\lambda)}\equiv\phi_D^{(2)}(\lambda).
\] 

{\bf Corollary 3.4} The  $M$-functions of the problems on $T_0$ with the roots at $v_1$ and $v_2$ are the same. This allows computer search for vertices on trees with equal $M$-functions to be quite fast. Previously it was necessary to attach intervals of arbitrary length to confirm that two vertices have the same $M$-function \cite{Pist} but we find here that for trees it is enough to attach intervals of length one.

{\bf Remark 3.5} If $d(v_1)\not=d(v_2)$ then co-spectrality of $T_1$ and $T_2$ does not imply $\phi_{1}(\lambda)\equiv\phi_{2}(\lambda)$.



\section{Examples}
 
1. Consider the graphs of Fig. 2. 

\begin{figure}[h]
 \begin{center}
   \includegraphics[scale= 0.6 ] {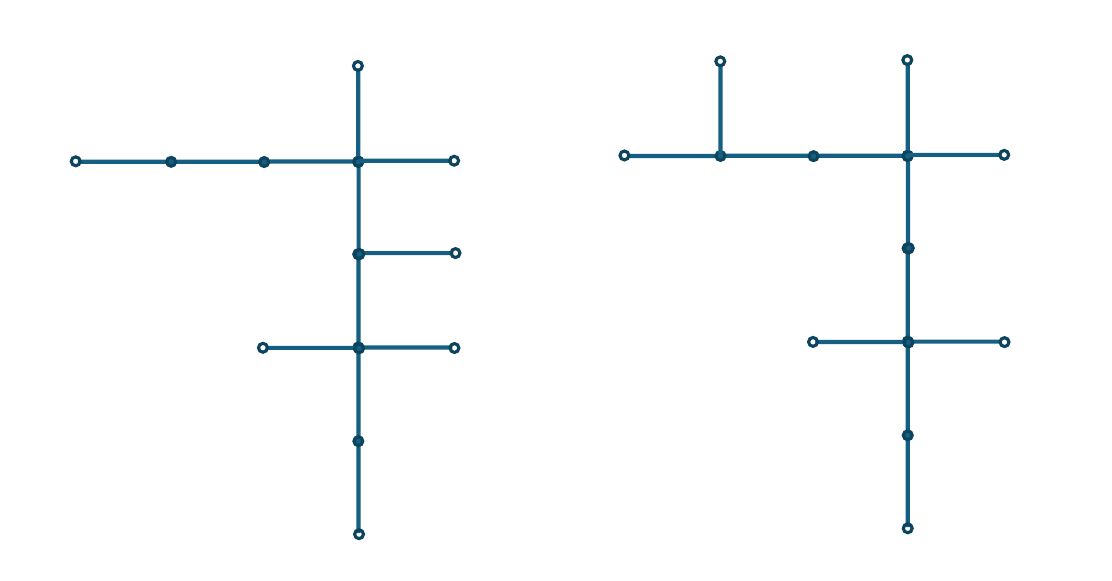}
  \end{center}
\caption{Trees obtained by attaching $P_2$ to the tree $T_0$ of Fig. 3 at the vertices $v_1$ and $v_2$, respectively. } 
\end{figure}

\begin{figure}[h]
 \begin{center}
   \includegraphics[scale= 0.6 ] {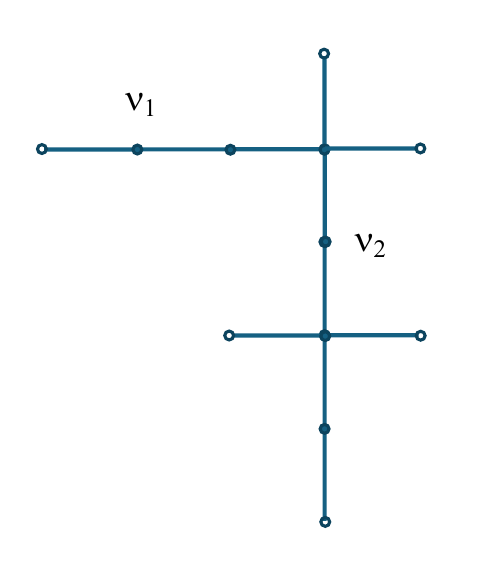}
  \end{center}
\caption{Tree $T_0$ corresponding to Examples 1 and 2.}
\end{figure}

They are obtained by attaching one edge to the tree of Fig. 3 at the vertices $v_1$ and $v_2$.  
Thus, in our terms ${\tilde T}$ is $P_2$ and $T_0$ is the tree of Fig. 3 and therefore 
\[
\phi_N^{(1)} (\lambda)=\phi_N^{(2)}(\lambda)=s^5(\lambda,l)(256c^6(\lambda,l)-192c^4(\lambda,l)+36c^2(\lambda,l)-1.
\]

The characteristic function of the Sturm-Liouville problem with the Dirichlet boundary conditions at the pendant vertices   on both graphs $T_1$ and $T_2$ is the same:

\[\phi_{1}(\lambda)=\phi_{2}(\lambda)=s^6(\lambda,l)(384c^6(\lambda,l)-256c^4(\lambda,l)+42c^2(\lambda,l)-1)\].




\begin{figure}
 \begin{center}
   \includegraphics[scale= 0.7 ] {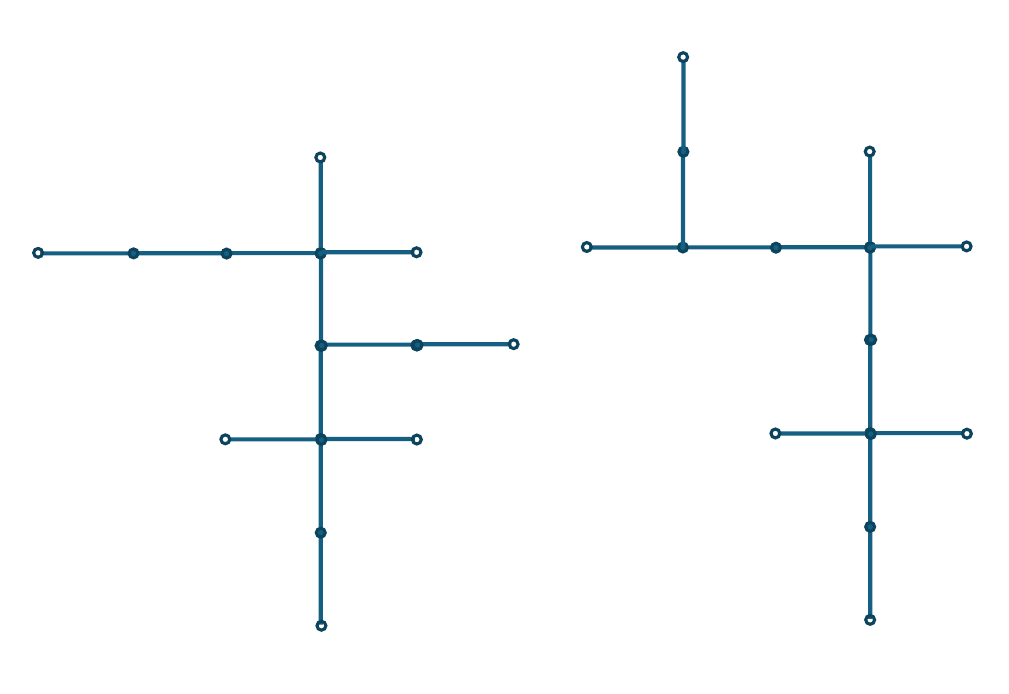}
  \end{center}
\caption{The tree $T_0$ of Fig. 3 with $P_3$ attached to the vertices $v_1$ and $v_2$, respectively.} 
\end{figure}

2. For the trees of Fig. 4 we have $\tilde{T}=P_3$ and $T_0$ is the tree of Fig. 3. Calculations show that 
\[
\phi_{1}(\lambda)\equiv{\phi}_{2}(\lambda)=-s^6(\lambda,l)(-768c^7(\lambda,l)+640c^5(\lambda,l)-148c^3(\lambda,l)+8c(\lambda,l)), 
\]
the same for both trees. 

3. Consider the trees presented in Fig. 34 b) of  \cite{Pist}, see Fig. 5.                                                                                                                                                                                                                                                                                                                                                                                                                                                                                    For these graphs $\phi_1(z)=(-24c^3(\lambda,l)+4c(\lambda,l)) s^5(\lambda,l)$, $\phi_2(z)=(-30c^3(\lambda,l)+5c(\lambda,l))s^5(\lambda,l)$. These trees can be obtained by attaching an edge (graph $P_2$ ) to the tree of Fig. 6 at $v_1$ and $v_2$, respectively. For both these trees $\phi_N^{(1)}(\lambda)\equiv\phi_N^{(2)}=(-20c^3(\lambda,l)+4c(\lambda,l)s^4(\lambda,l)$. For the first tree $\phi^{(1)}_D=4c^2(\lambda,l)s^5(\lambda,l)$ and and for the second tree $\phi_D^{(2)}(\lambda)=(10c^2(\lambda,l)-1)s^5(\lambda,l)$.  The reason of this difference  is coused by thedifference in degrees $d(v_1)=5\not=2=d(v_2)$.
\begin{figure}
\begin{center}
\includegraphics[scale= 0.7 ] {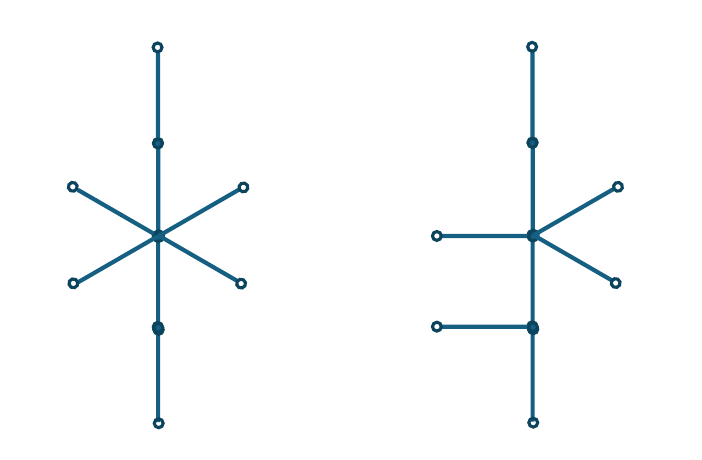}
\end{center}
\caption{The trees obtained by attaching $P_2$ to the vertices $v_1$ and $v_2$ to the tree of Fig. 6.}
\end{figure}

\begin{figure}
 \begin{center}
   \includegraphics[scale= 0.7 ] {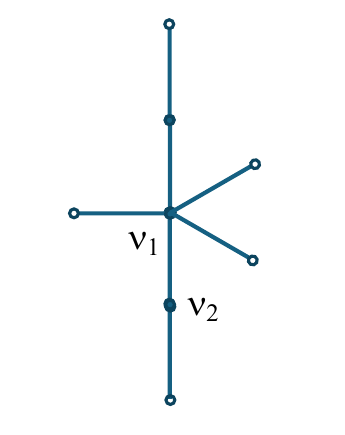}
  \end{center}
\caption{The tree $T_0$ of Example 3.}
\end{figure}
\newpage
{\bf Acknowledgements}

The present research was supported by the Academy of Finland (project no. 358155). V.P. is grateful to the University of Vaasa for hospitality.
V.P. is grateful to the Ministry of Education and Science of Ukraine for the support in completing the work 'Inverse problems of finding the shape of a graph by spectral data' State registration number 0124U000818 and to NSF US for support IMPRESS-U: Spectral and geometric methods for damped wave equations with applications to fiber lasers.

$^1$ Mats-Erik Pistol  Solid State Physics, Lund University, S-221 00 Lund, Sweden 
mats-erik.pistol@ftf.lth.se

$^2$ Vyacheslav Pivovarchik South Ukrainian National Pedagogical University (Odesa, Ukraine0 and University of Vaasa (Vaasa, Finland) vpivovarchik@gmail.com

\end{document}